\documentclass[aps,twocolumn,superscriptaddress,amsmath,amssymb,pra]{revtex4-2}
\usepackage{graphicx}% Include figure files
\usepackage{dcolumn}% Align table columns on decimal point
\usepackage{bm}% bold math
\usepackage{braket}
\usepackage{appendix}
\usepackage{subfigure}
\usepackage{hyperref}
\usepackage{color}
\usepackage{xcolor}
\usepackage{amsmath}
\usepackage{nccmath}
\usepackage[english]{babel}
\usepackage{svg}
\begin{document}

\title[time crystalline order]{Stability of the discrete time-crystalline order in spin-optomechanical and open cavity QED systems}

\author{Zhengda Hu} 
\affiliation{Department of Physics and Astronomy, Purdue University, West Lafayette, Indiana 47907, USA}
\affiliation{School of Science, Jiangnan University, Wuxi 214122, China}

\author{Xingyu Gao} 
\affiliation{Department of Physics and Astronomy, Purdue University, West Lafayette, Indiana 47907, USA}
\affiliation{School of Science, Jiangnan University, Wuxi 214122, China}

\author{Tongcang Li}
\email{tcli@purdue.edu}
\affiliation{Department of Physics and Astronomy, Purdue University, West Lafayette, Indiana 47907, USA}
\affiliation{Elmore Family School of Electrical and Computer Engineering, Purdue University, West Lafayette, Indiana 47907, USA}
\affiliation
{Birck Nanotechnology Center, Purdue University, West Lafayette,
	IN 47907, USA}
\affiliation
{Purdue Quantum Science and Engineering Institute, Purdue University, West Lafayette, Indiana 47907, USA}

\date{\today}

\begin{abstract}
Discrete time crystals (DTC) have been demonstrated experimentally in several different quantum systems in the past few years. Spin couplings and cavity losses have been shown to play crucial roles for realizing DTC order in open many-body systems out of equilibrium. Recently, it has been proposed that eternal and transient DTC can be present with an open Floquet setup in the thermodynamic limit and in the deep quantum regime with few qubits, respectively.
In this work, we consider the effects of spin damping and spin dephasing on the DTC order in spin-optomechanical and open cavity systems in which the spins can be all-to-all coupled.
In the thermodynamic limit, it is shown that the existence of dephasing can destroy the coherence of the system and finally lead the system to its trivial steady state.
Without dephasing, eternal DTC is displayed in the weak damping regime, which may be destroyed by increasing the all-to-all spin coupling or the spin damping.
By contrast, the all-to-all coupling is constructive to the DTC in the moderate damping regime. We also focus on a model which can be experimentally realized by a suspended hexagonal boron nitride (hBN)  membrane with a few spin color centers under microwave drive and Floquet magnetic field. Signatures of transient DTC behavior are demonstrated in both weak and moderate dissipation regimes without spin dephasing.
Relevant experimental parameters are also discussed for realizing transient DTC order in such an hBN optomechanical system.
\end{abstract}
%\keyword{discrete time-crystalline order; decoherence; optomechanical systems; hexagonal boron nitride}

\maketitle

%%%%%%%%%%%%%%%%%%%%%%%%%%%%%%%%%%%%%%%%%%
%\setcounter{section}{-1} %% Remove this when starting to work on the template.
\section{Introduction}
In recent years, periodically driven (Floquet) quantum many-body systems have attracted considerable attention since they are crucial for understanding new non-equilibrium Floquet many-body localization (MBL)~\cite{Abanin2019}  phase and may have potential applications in quantum metrology ~\cite{Lyu2020}.
One example of a non-equilibrium Floquet-MBL phase is the discrete time-crystalline (DTC) order~\cite{Sacha2015,Else2016,Khemani2016},
which is different from a continuous time crystal~\cite{Wilczek2012,Li2012,Huang2018,Huang2020} and is characterized by the breaking of discrete time-translation symmetry (TTS)~\cite{Sacha2020}. The DTC order has been realized experimentally in several  quantum systems in the past few years~\cite{Choi2017,Zhang2017,Randall2021,Kyprianidis2021}. Under driving with a period $T$, the system can exhibit stroboscopic response with a period $nT$ and it is expected to be robust against imperfection of the driving~\cite{von Keyserlingk2016,Yao2017}.
Recently, the DTCs in open Floquet systems have been \mbox{reported~\cite{Lazarides2017,Else2017,Gong2018,Zhu2019,KL2021,Lazarides2020,Riera-Campeny2020}.}
Since any realistic systems will be unavoidably coupled to its surroundings and the influences of baths can be either negative or positive, the mechanisms of stabilizing DTC in dissipative systems will be important to explore.

\begin{figure}[tbh!]
	%\centering
	\includegraphics[width=0.47\textwidth]{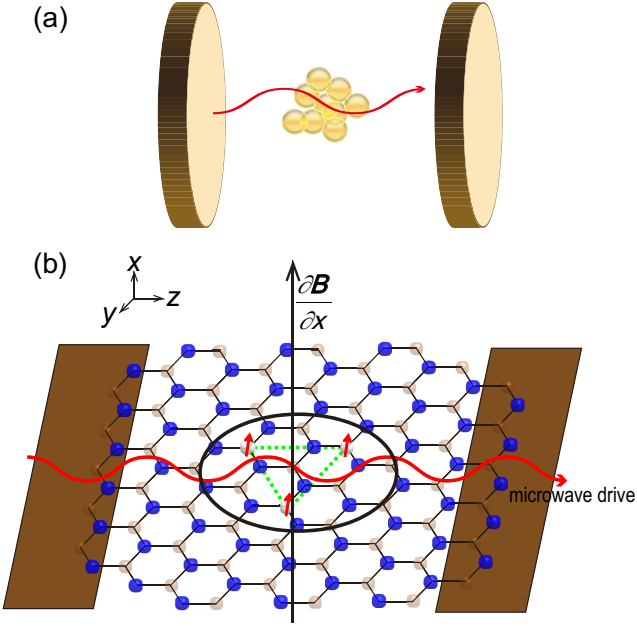}
	\caption{Sketches of the setups for realizing the DTC order: ({\bf a}) a large ensemble of spins trapped in a cavity and ({\bf b}) a suspended hBN monolayer membrane with a few spin color centers under a microwave drive and a circular-localized magnetic field.  The Hamiltonian is modeled by Equation~(\ref{H}) and the Floquet driving protocol is that the spin-cavity coupling $\lambda$ is switched on (off) in the first (second) half of a Floquet period $T$. In this work, both spin and cavity losses are considered.}
	\label{fig:fig1}
\end{figure}

Meanwhile, recent development of  optomechanical systems~\cite{Fabre1994,Mancini1994,RMP2014,Yin2015,Xu2021casimir} has facilitated breakthroughs of quantum technologies such as ground state cooling~\cite{Chan2011,Liu2013}, optical sensing~\cite{Xiong2017,Liu2017,Krause2012,Ahn2020, LiBB2021}, and quantum information processing~\cite{Stannigel2010,Stannigel2012}.
With nanoscale cavity optomechanical devices, the coupling between light and motion of mechanical resonators can be flexibly modulated with controllable loss~\cite{Karg2020}, which may even reach ultrastrong coupling regime~\cite{Frisk Kockum2019}.
A natural choice of mechanical modes is to use membranes of two-dimensional materials due to their excellent mechanical properties~\cite{Akinwande2017}.
Recently, hexagonal boron nitride (hBN) has drawn great interest and served as a promising platform for exploring both quantum and nanophotonic effects~\cite{Tran2016,Cadiz2018,Liu2019,Xia2014,Klusek2010}. hBN has a very wide bandgap and outstanding chemical and thermal stability beyond that of graphene. As a type of van der Waals materials, hBN can be integrated with plasmonic, nanophotonic, and potentially more complex structures~\cite{Tran2017,Caldwell2019,Gao2020,Wu2021,Xu2021}. The hBN membranes have low mass, small out-of-plane stiffness, high elasticity modulus and strong tensile strength, which make them a promising candidate for high-Q mechanical resonators and high-sensitivity sensors~\cite{Kim2018,Shandilya2019}.
A spin-mechanical system based on color centers in a suspended hBN mechanical resonator has been proposed~\cite{Abdi2017,Abdi2019},
which can even simulate the Rabi model in the ultrastrong coupling regime. Very recently, optically addressable spin defects were observed in hBN~\cite{Gottscholl2020,Chejanovsky2021,Gao2021}. As the DTC order has been found in $N$ atoms in a lossy cavity ~\cite{Gong2018,Zhu2019,KL2021}, it is interesting to explore the DTC in such spin-optomechanical systems with incoherent noise (spin damping or~dephasing).

In this work, we consider the DTC behaviors in an open Floquet system as $N$ qubits in a (mechanical) cavity via switching on and off of the spin-cavity coupling. In the thermodynamic limit, it describes a cavity QED model with a large ensemble of trapped spins while, in the deep quantum regime (with few qubits), it characterizes an optomechanical model as a suspended hBN monolayer membrane with a few spin defects under a microwave drive and a Floquet magnetic field (Figure \ref{fig:fig1}). We discuss stroboscopic dynamics in both regimes and explore whether stroboscopic oscillations are stable to spin damping and spin dephasing as well as the effect of all-to-all spin coupling.

\section{Perfect DTC in the Thermodynamic Limit}

We consider an open system as $N$ qubits with all-to-all interactions in a (mechanical) cavity (Figure \ref{fig:fig1}).  The all-to-all coupling can be mediated by a photon in an optical cavity~\cite{Gong2018} or a phonon in a mechanical oscillator~\cite{Abdi2017,Abdi2019,LiB2020}. The Hamiltonian is given by~\cite{Gong2018,Zhu2019,Abdi2017,Abdi2019,Morrison2008,Russomanno2017}
\begin{equation}\label{H}
	\hat{H}(h,\lambda)=\omega_{0}\sum_{i}\hat{s}_{i}^{z}+\omega\hat{a}^{\dag}%
	\hat{a}+\frac{2h}{N}\sum_{i<j}\hat{s}_{i}^{z}\hat{s}_{j}^{z}+\frac{2\lambda
	}{\sqrt{N}}(\hat{a}+\hat{a}^{\dag})\sum_{i}\hat{s}_{i}^{x},
\end{equation}
where $\hat{a}$ ($\hat{a}^{\dag}$) is the annihilation (creation) operator of the photon field with optical frequency $\omega$,
$\hat{s}_{i}^{\mu}$ ($\mu=x,y,z$) is the spin-$\frac{1}{2}$ angular momentum operator along the $\mu$ axis for the $i$-th qubit
of transition frequency $\omega_{0}$, and $h$ ($\lambda$) is related to the spin-spin (spin-cavity) coupling strength.
For convenience, a more compact version can be derived as
\begin{equation}
	\hat{H}(h,\lambda)=\omega_{0}\hat{J}_{z}+\omega\hat{a}^{\dag}\hat{a}+\frac
	{h}{N}\hat{J}_{z}^{2}+\frac{2\lambda}{\sqrt{N}}(\hat{a}+\hat{a}^{\dag})\hat
	{J}_{x},
\end{equation}
by introducing the collective angular moment operator $\hat{J}_{\mu}=\sum_{i}\hat{s}_{i}^{\mu}$ and neglecting a constant term.
We consider a general decoherent model by including both the spin and cavity losses.
Then, the dynamics of the system can be described by the master equation (setting $\hbar=1$)
\begin{equation}\label{mseq}
	\frac{\mathrm{d}\hat{\rho}}{\mathrm{d}t}=-\mathrm{i}[\hat{H},\hat{\rho
	}]+\gamma D[\hat{a}]\hat{\rho}+\frac{\Gamma}{N}D[\hat{J}_{-}]\hat{\rho}%
	+\frac{\tilde{\Gamma}}{N}D[2\hat{J}_{z}]\hat{\rho},
\end{equation}
where $\hat{J}_{-}=\hat{J}_{x}-\mathrm{i}\hat{J}_{y}$ is the collective lowering operator and
$D[\hat{o}]\hat{\rho}=\hat{o}\hat{\rho}\hat{o}^{\dag}-(\hat{o}^{\dag}\hat{o}\hat{\rho}+\hat{\rho}\hat{o}^{\dag}\hat{o})/2$. Here, $\gamma=\omega/Q$
is the cavity damping rate with $Q$ the quality factor.
In addition, $\Gamma$ and $\tilde{\Gamma}$ are the spin relaxation and dephasing rate, respectively. Previous works mainly focused on the DTC in cavity QED systems with merely the cavity loss or the nearest-neighbor (short-range) spin coupling~\cite{Gong2018,Zhu2019,KL2021}. They have neither discussed stabilizing DTC in dissipative systems with all-to-all coupling nor considered the effects of spin damping and spin dephasing.

%\end{paracol}
%\nointerlineskip
\begin{figure*}[tbh!]
	\centering
	\includegraphics[angle=0,width=15cm]{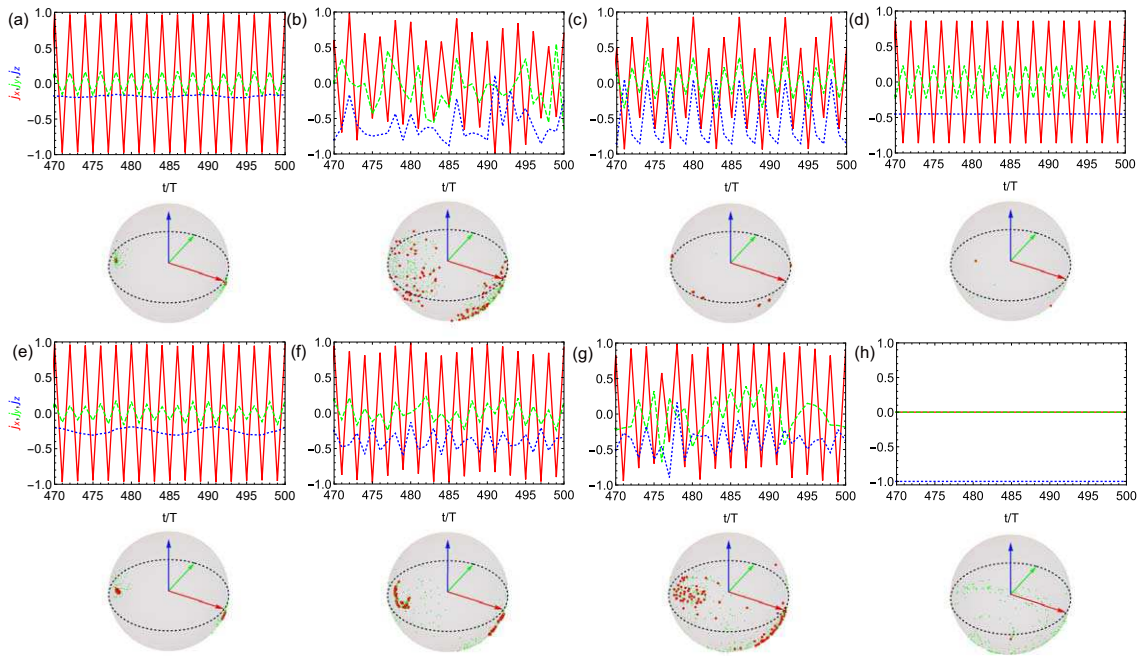}
	\caption{Stroboscopic dynamics (\textbf{top}) and stroboscopic trajectories (\textbf{bottom}) of the scaled angular momentum vector $\vec{j}=(j_x,j_y,j_z)$ (red, green, blue) in the thermodynamic limit for the perfect driving case $\varepsilon=0$. The top shows typical stroboscopic dynamics of $j_x$ (solid red curve), $j_y$ (dashed green curve) and $j_z$ (dotted blue curve) for the last $30$ periods of the entire $500$-period evolution. The bottom displays the stroboscopic trajectories on the Bloch sphere for the entire $500$ periods (green) and for the last $100$ periods (red) sphere. We consider no spin-cavity coupling $h=0$ in ({\bf a}--{\bf d}) and  increasing spin-cavity coupling strength in ({\bf e}--{\bf h}) with $h=0.05$, $0.1$, $0.3$, $1$, respectively. The parameters are set as: ({\bf a},{\bf e}) $\gamma=\Gamma=0.05$, ({\bf b},{\bf f}) $\gamma=0.05$, $\Gamma=0.3$, (\textbf{c},\textbf{g}) $\gamma=\Gamma=0.3$ and (\textbf{d},\textbf{h}) $\gamma=1.5$, $\Gamma=0.3$.  }
	\label{fig:fig2}
\end{figure*}

First, we would like to consider the robustness of DTC behavior in the thermodynamic limit $N\rightarrow\infty$.
By performing the mean-field approximation and factorizing the means of operator product, we obtain a closed set of semiclassical equations as
\begin{align}\label{mfeqns}
	\dot{j}_{x}  &  =-\omega_{0}j_{y}-hj_{y}j_{z}+\frac{\Gamma}{2}j_{x}%
	j_{z}-\tilde{\Gamma}j_{x},\nonumber\\
	\dot{j}_{y}  &  =\omega_{0}j_{x}-2\lambda\sqrt{2\omega}xj_{z}+hj_{x}%
	j_{z}+\frac{\Gamma}{2}j_{y}j_{z}-\tilde{\Gamma}j_{y},\nonumber\\
	\dot{j}_{z}  &  =2\lambda\sqrt{2\omega}xj_{y}+\frac{\Gamma}{2}(j_{z}^{2}-1),\nonumber\\
	\dot{x}  &  =p-\frac{\gamma}{2}x,\nonumber\\
	\dot{p}  &  =-\omega^{2}x-\frac{\gamma}{2}p-2\lambda\sqrt{2\omega}j_{x},
\end{align}
where $j_{\mu}= \langle \hat{J}_{\mu} \rangle /j$ with $j=N/2$ and ${\sum_{\mu}}j_{\mu}^{2}=1$, $x=\langle \hat{a}+\hat{a}^{\dag}\rangle/\sqrt{2N\omega}$, and $p=\mathrm{i}\langle \hat{a}^{\dag}-\hat{a}\rangle /\sqrt{2N/\omega}$.
The set of Equation~(\ref{mfeqns}) is a generalization of that in Reference~\cite{Gong2018} which is a special case as $h=0$ here.
The introduction of spin-spin coupling $h$ breaks the original stable attractors $(j_x, j_y, j_z)_\mathrm{st}=(\pm \sqrt{1-\mu^2},0,-\mu)/2$ and $(x,p)_\mathrm{st}=\mp[\lambda\sqrt{2\omega(1-\mu^2)}/(\omega^2+\gamma^{2}/4)](1,\gamma/2)$,
with $\mu=(\lambda_\mathrm{c}/\lambda)^2$ and the critical spin-cavity coupling strength $\lambda_\mathrm{c}=\sqrt{(\omega_0/\omega)(\omega^2+\gamma^2/4)}/2$.
We would also like to focus on the steady-state solutions as Reference~\cite{Gong2018}, which is instead numerically found out due to the more complexity considered.
It is clear that there exist trivial steady-state solutions as $x=p=j_{x}=j_{y}=0$ and $j_{z}=\pm1$.
Besides, as long as the dephasing exists ($\tilde{\Gamma}\neq0$), the steady-state solutions will fall into be trivial.
This can be understood as that the existence of dephasing will finally destroy the coherence (non-diagonal terms of density matrix)
and leads to the final state as either $\left\vert +N/2\right\rangle$ or $\left\vert -N/2\right\rangle$ when the $\mathbb{Z}_2$ symmetry is broken at $\lambda>\lambda_\mathrm{c}$. Here, $\left\vert \pm N/2\right\rangle $ are the eigenstates of $\hat{J}_{z}$ with
$\hat{J}_{z}$ $\left\vert \pm N/2\right\rangle =\pm N/2\left\vert \pm N/2\right\rangle$.
Therefore, we set $\tilde{\Gamma}=0$ in the following discuss, unless specifically mentioned.
Besides, we assume the spins are initially in the eigenstate $\left\vert \rightarrow\rightarrow\cdots\rightarrow\rightarrow
\right\rangle $ with $j_{x}|_{t=0}=1$, $j_{y}|_{t=0}=0$, and $j_{z}|_{t=0}=0$ and the cavity mode is initially in a coherent state $\left\vert
\alpha\right\rangle $ with $x|_{t=0}=p|_{t=0}=0$. If we consider the symmetry-broken regime $\lambda>\lambda_\mathrm{c}$, it is clear that the final state will fall into either one of the two nontrivial stable states. To observe a DTC order, we perform the Floquet driving protocol similar to Reference~\cite{Gong2018}: the spin-cavity coupling $\lambda$ is artificially switched off in the second-half period, i.e., $\lambda=0$ for $(n+1/2)T \leq t<(n+1)T$ with $n=0,1,2,\ldots$. From an alternative viewpoint, the Floquet driving is that we let the spins periodically driven by a leaky cavity in every first-half period $nT \leq t<(n+1/2)T$. We introduce the imperfection parameter $\varepsilon$ via a detuning between $\omega$ and $\omega_0$ as $\omega=(1-\varepsilon)\omega_\mathrm{T}$ and $\omega_0=(1+\varepsilon)\omega_\mathrm{T}$ with $\omega_\mathrm{T}=2\pi/T$. In the perfect case ($\varepsilon=0$), it is not difficult to check that the unitary dynamics during the second-half period contributes a parity operator $P=\mathrm{e}^{-\mathrm{i}\pi(a^{\dag}a+J_z)}$ which flips the stable state to the other one. If certain observables of the spins (say $j_\mu$) or the cavity mode (say $x,p$) exhibit period doubling oscillations which are robust against imperfection driving $\varepsilon$, then a DTC order may be identified. We also consider nonunitary imperfections due to decoherence of the system. To observe the long-time behavior, we numerically solve a Floquet--Lindblad master equation (setting $\lambda$ in Equation~(\ref{mfeqns}) be periodically time-dependent as characterized above) up to $500$ periods $T$ by means of the Runge-Kutta method. We shall remark that we have also tried more periods such as $5000$ periods as in Reference~\cite{Gong2018} but there is no qualitative difference. For convenience, we set $\omega_\mathrm{T}=1$ and $\lambda=1$ to illustrate the perfect DTC in the $\lambda>\lambda_\mathrm{c}$ regime.

\begin{figure*}[tbh!]
	\centering
	\includegraphics[angle=0,width=15cm]{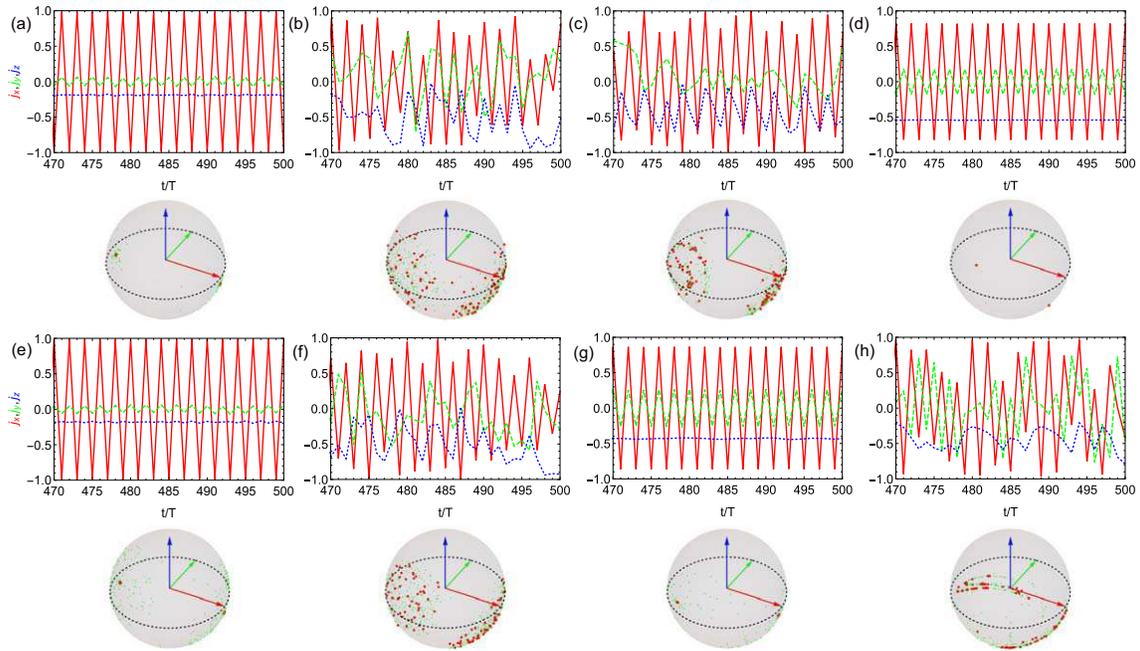}
	\caption{Stroboscopic dynamics 
		(\textbf{top}) and stroboscopic trajectories \textbf{(bottom}) of the scaled angular momentum vector components $j_x$ (solid red curve), $j_y$ (dashed green curve), $j_z$ (dotted blue curve) in the thermodynamic limit for the imperfect driving case $\varepsilon=0.05$. Other parameters are the same as Figure~\ref{fig:fig2}. The robustness of DTC against imperfection is clearly shown in ({\bf a},{\bf d},{\bf e}). Besides, it is interesting to find that the DTC can even benefit from the imperfection as comparing Figure~\ref{fig:fig3}g with Figure~\ref{fig:fig2}g.}
	\label{fig:fig3}
\end{figure*}

In Figures~\ref{fig:fig2} and~\ref{fig:fig3}, we plot the stroboscopic dynamics of the scaled angular momentum vector $\vec{j}=(j_x,j_y,j_z)$ as well as their stroboscopic trajectories on the Bloch sphere for the perfect driving ($\varepsilon=0$) and imperfect driving ($\varepsilon\neq 0$) cases, respectively.
By comparing the first row a-d where there is no spin-spin coupling with $h=0$, we clearly observe different stroboscopic dynamics in different dissipation regimes. First, the DTC order is well preserved by the existence of weak spin damping $\Gamma$ as shown in Figure~\ref{fig:fig2}a and robust again imperfection $\varepsilon$ as shown in Figure~\ref{fig:fig3}a. As the spin damping rate $\Gamma$ increases, the DTC dynamics becomes irregular with the trajectory of $\vec{j}$ scattered on the Bloch sphere (Figures~\ref{fig:fig2}b and ~\ref{fig:fig3}b). However, the dynamics will become more regularly with the area of stroboscopic trajectories reduced if the cavity loss rate $\gamma$ increases (Figures~\ref{fig:fig2}c and ~\ref{fig:fig3}c). The eternal stroboscopic oscillations will occur again with the trajectories almost collapse into the two stable points for $\gamma\gg \Gamma$ (Figure~\ref{fig:fig2}d), which is robust against imperfection $\varepsilon$ (Figure~\ref{fig:fig3}d) so as to identify the DTC order.
Besides, by comparing the second row (e-h) where there is spin-spin coupling $h\neq 0$,  different stroboscopic dynamics from that of $h=0$ is also demonstrated in different dissipation regimes. From Figures~\ref{fig:fig2}e--h (perfect $\varepsilon=0$ case) with growing all-to-all coupling $h$, we observe that DTC oscillations is gradually destroyed and the system finally falls into one of the trivial stable states with $j_x=j_y=0$ and $j_z=-1$ (Figure~\ref{fig:fig2}h).
By contrast, in the imperfect case ($\varepsilon\neq 0$) as shown in Figures~\ref{fig:fig3}e--h, we surprisingly find that the DTC order may be rebuilt by appropriate $h$ in the moderate damping regime, by comparing Figure~\ref{fig:fig3}g with Figure~\ref{fig:fig2}g.

%\begin{paracol}{2}
%\linenumbers
%\switchcolumn

\section{Transient DTC Behavior in the Deep Quantum Regime}

We proceed to focus on the few-atom cases [$N\sim O(1)$], which corresponds to the hBN optomechanical system as displayed in Figure~\ref{fig:fig1}b.
It is expected that a DTC behavior may still survive in few atom cases, the so-called deep quantum regime~\cite{Gong2018}.
In this regime, we do not perform semiclassical approximation so that all the quantumness of the system is well maintained.
The interplay among spin-spin coupling, spin-cavity coupling and dissipations may give rise to more subtle behaviors for transiently long DTC
in this deep quantum regime. By transiently long we mean that the DTC lasts much longer than the decay time $\gamma^{-1}$. The initial state is chosen to be $\left\vert \Rightarrow \right\rangle $ $\otimes$ $\left\vert \alpha\right\rangle $,
where $\left\vert\Rightarrow\right\rangle \equiv\otimes_{j=1}^{N}\left\vert \rightarrow\right\rangle $ is the eigenstate of $\hat{J}_{x}$ with the eigenvalue $N/2$ and $\left\vert \alpha\right\rangle $ is a coherent state with $\hat{a}\left\vert \alpha\right\rangle =\alpha\left\vert \alpha\right\rangle $.
The Floquet--Lindblad dynamics extended from Equations~(\ref{H}) and (\ref{mseq}) is directly solved under a truncation of $16$ photons for $\alpha=0.01$.

\begin{figure*}[tbh!]
	\centering
	\includegraphics[angle=0,width=6cm]{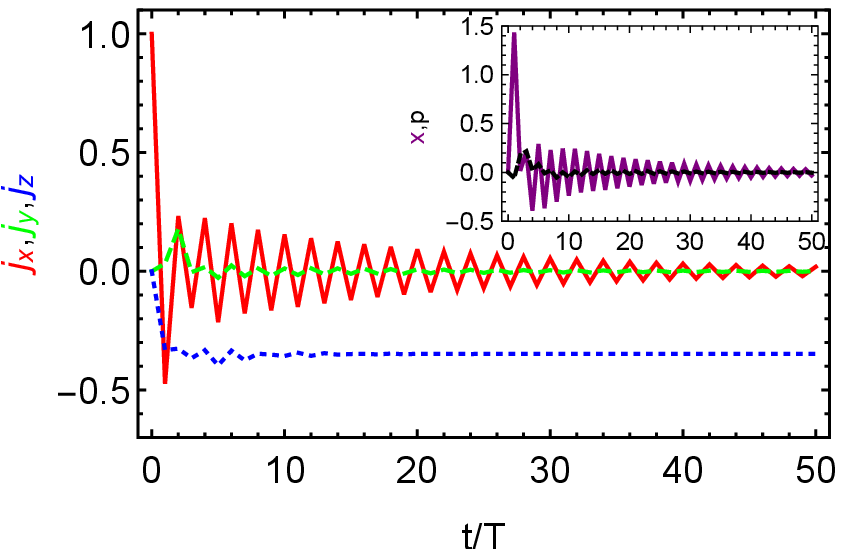}
	\includegraphics[angle=0,width=6cm]{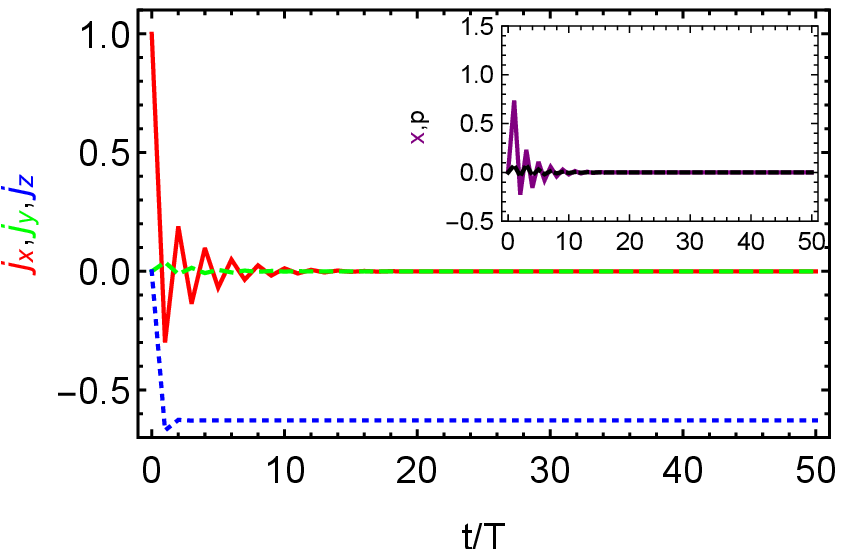}
	\includegraphics[angle=0,width=6cm]{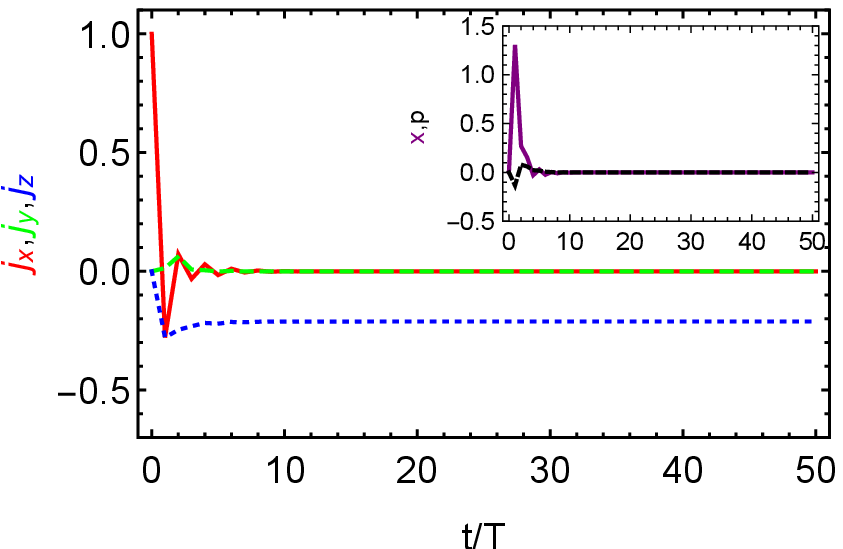}
	\includegraphics[angle=0,width=6cm]{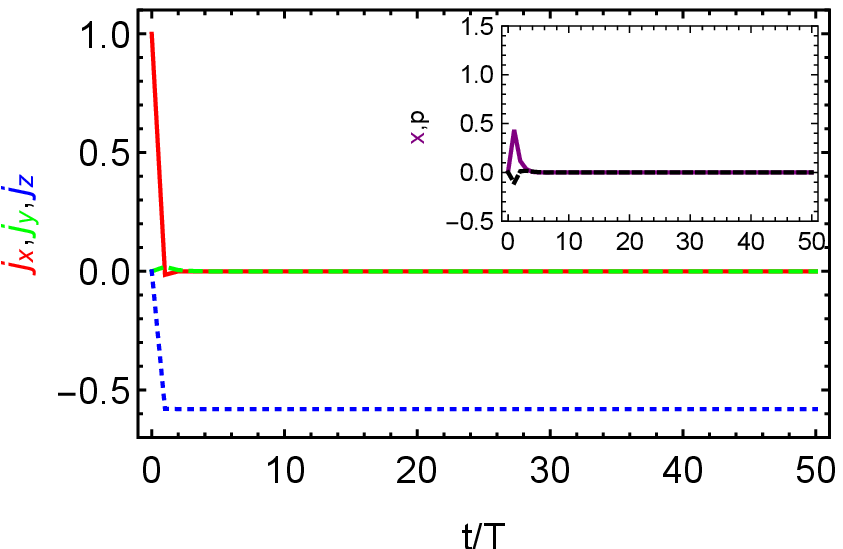}
	\caption{Stroboscopic %MDPI: Please check if the copyright permission is needed for Figure 4. If so, please provide it.
		dissipative dynamics of the scaled angular momenta of $j_x$ (red solid curve), $j_y$ (green dashed curve), and $j_z$ (blue dotted curve) in the two-qubit $N=2$ case. The inset shows quadratures $x$ (purple solid) and $p$ (black dashed) behaviors. We consider weak  dissipation in (\textbf{a}) $h=\gamma=\Gamma=0.05, \tilde{\Gamma}=0$ and moderate  dissipation in (\textbf{b}) $h=\gamma=\Gamma=0.3, \tilde{\Gamma}=0$ but without spin dephasing as the thermodynamic limit case. Contrast to (\textbf{a}) and (\textbf{b}), (\textbf{c}) and (\textbf{d}) includes  spin dephasing $\tilde{\Gamma}\approx 2\Gamma$ as suggested in Reference~\cite{Abdi2017}.}
	\label{fig:fig4}
\end{figure*}

\begin{figure*}[tbh!]
	\centering
	\includegraphics[angle=0,width=6cm]{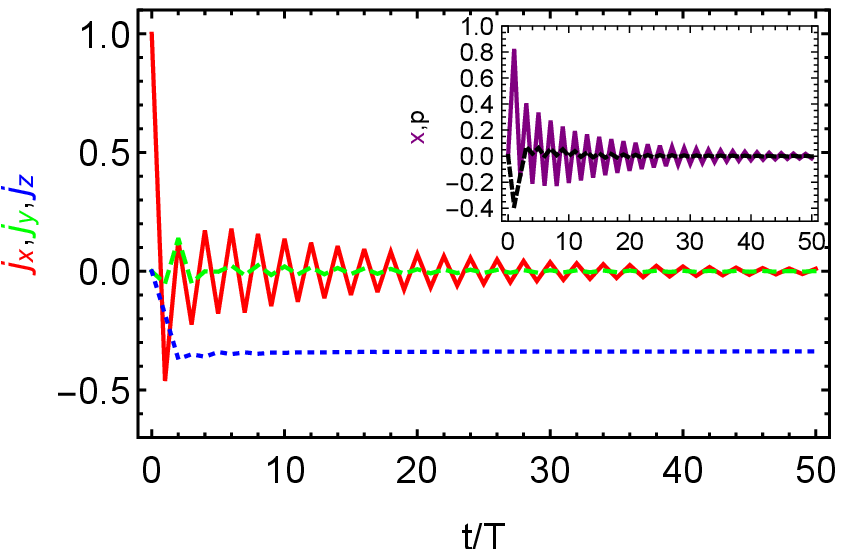}
	\includegraphics[angle=0,width=6cm]{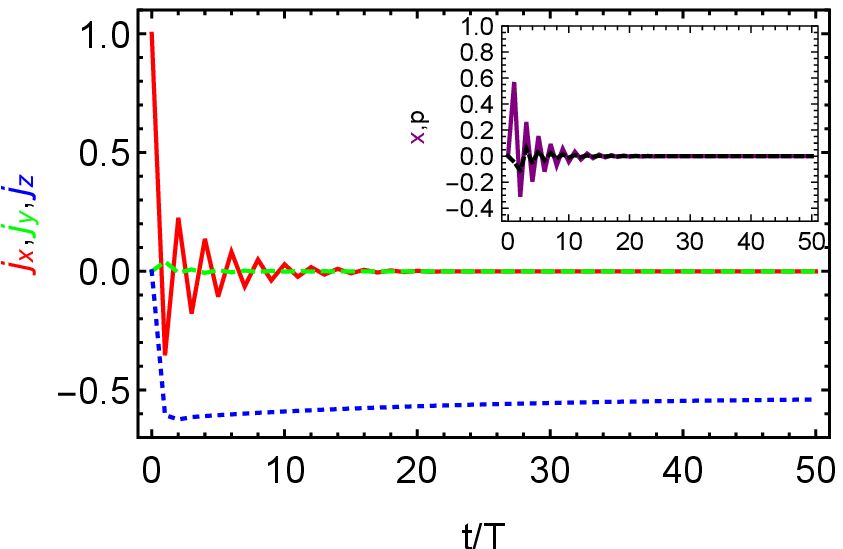}
	\includegraphics[angle=0,width=6cm]{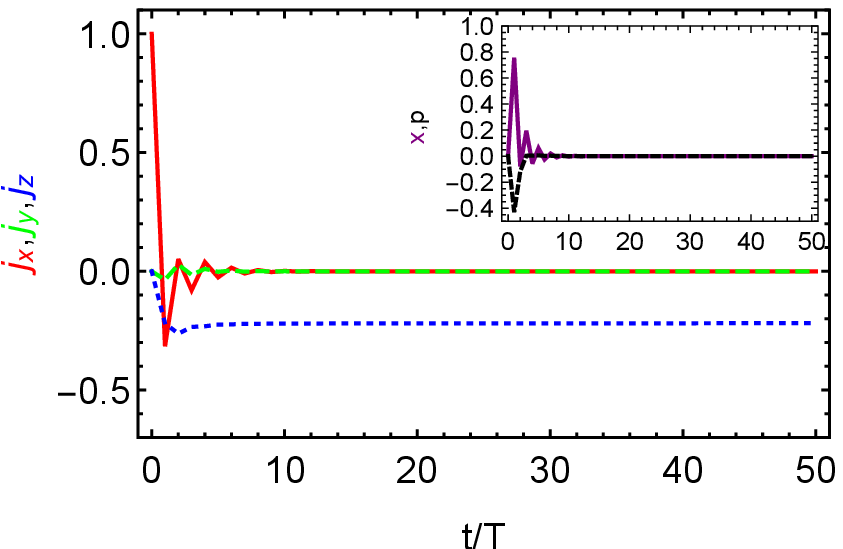}
	\includegraphics[angle=0,width=6cm]{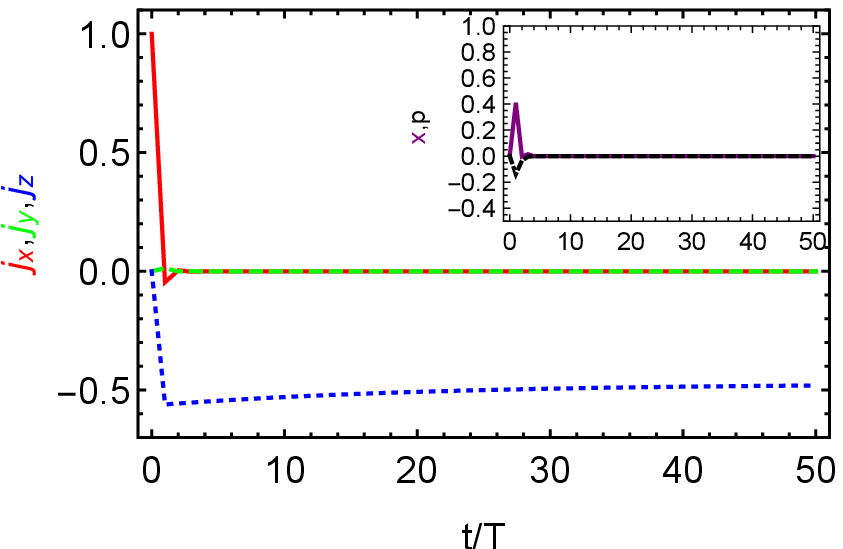}
	\caption{Stroboscopic 
		dissipative dynamics of the scaled angular momenta of $j_x$ (red solid curve), $j_y$ (green dashed curve), and $j_z$ (blue dotted curve) for the
		three-qubit $N=3$ case. The parameter setups are the same as those in Figure~\ref{fig:fig5}.}
	\label{fig:fig5}
\end{figure*}

Figure \ref{fig:fig4}a shows the  stroboscopic dynamics of the scaled angular momenta $j_{\mu}$ and quadratures $x$, $p$ (inset)
in the strong coupling regime ($\lambda=1$)
and weak dissipation regime ($\gamma=\Gamma=0.05$) for the two-qubit case ($N=2$).
We clearly observe that $j_{x}$ and $x$ exhibit stroboscopic oscillations with doubling period $2T$ after $t\sim5T$,
which persists even at $t\sim50T$ and thus is much longer than the decay time (here $\gamma^{-1}=\Gamma^{-1}\sim3T$). In this sense, a transient DTC order is established in the deep quantum regime before reaching the stationary state.
In Figure~\ref{fig:fig4}b, we plot the  stroboscopic dynamics in the moderate dissipation regime ($\gamma=\Gamma=0.3$). In this case, the decay time
can be estimated as $\gamma^{-1}=\Gamma^{-1}\sim 0.5 T$ so that the stroboscopic dynamics occur immediately and lasts over $10 T$,
which still maintains a transient DTC order.
Moreover, if the spin dephasing $\tilde{\Gamma}\approx 2 \Gamma$ as predicted in Reference~\cite{Abdi2017} is additionally considered, as shown in Figures~\ref{fig:fig4}c,d, we find that the oscillation time is merely comparable to the decay time and thus no transient DTC order exists. Besides, we observe similar phenomena if more spins are involved such as the case of $N=3$ shown in Figure~\ref{fig:fig5}. One effect of increasing the spin number $N$ is that the transient oscillations evolve into an eternal one as predicted at the thermodynamic limit $N\rightarrow \infty$ in Figure~\ref{fig:fig2}. Another effect of increasing $N$ may be that the stroboscopic oscillations is more robust against the spin dephasing as comparing the oscillation dynamics of quadrature $x$ (purple solid) in Figures~\ref{fig:fig4}c and ~\ref{fig:fig5}c.

Before ending, we would like to discuss the setup of experimental parameters for realizing TDC order in the optomechanical system of hBN monolayer membrane. According to Referecne~\cite{Abdi2017}, a maximum magnetic field gradient $270$ $\mathrm{G}/\mathrm{nm}$ may be reached such that the spin-cavity coupling $\lambda$ may become comparable or even larger than the oscillator frequency $\omega$. In this work, we consider $\lambda=\omega_{\mathrm{T}}$, $\omega=(1-\varepsilon)\omega_{\mathrm{T}}$ and $\omega_0=(1+\varepsilon)\omega_{\mathrm{T}}$ with $\varepsilon\leq 10 \%$, and the cavity loss rate $\gamma\leq 1.5 \omega_{\mathrm{T}}$, which indicates $\lambda_\mathrm{c}\leq 0.65 \omega_{\mathrm{T}}$.
To insure the occurrence of transient DTC dynamics, we need operate in the regime of $\lambda >\lambda_\mathrm{c}$. Then, the minimal spin-cavity coupling is to achieve $\lambda >0.65 \omega_{\mathrm{T}}$, which is realizable in a suspended circular hBN membrane with radius $R\sim 1$ $ \mu \mathrm{m}$. Another important aspect is to control the dephasing rate $\tilde{\Gamma}$ which is detrimental to the DTC order. According to Reference~\cite{Abdi2017}, the spin dephasing mainly stems from optical polarization $\tilde{\Gamma}_\mathrm{o}$ and membrane vibrations $\tilde{\Gamma}_\mathrm{v}$, which is proportional to the vibration frequency $\omega$. Therefore, to suppress the dephasing rate, it is suggested to reduce the cavity frequency $\omega$, which also corresponds to enhance the membrane radius $R$. Last, but not least, the cavity loss promotes spin cooling and localization, which is crucial to the emergence of DTC. However, as can be indicated by comparing Figure ~\ref{fig:fig4}b with Figure ~\ref{fig:fig4}a (or Figure ~\ref{fig:fig5}b with Figure ~\ref{fig:fig5}a), a too strong cavity loss $\gamma$ (corresponding to extremely low $Q$) may overdamp the system dynamics and destroy the DTC order. Besides, a stronger $\gamma$ leads to a higher critical spin-cavity coupling $\lambda_\mathrm{c}$ such that stronger spin-cavity coupling $\lambda$ is needed, which imposes a challenge to its experimental realization. For cavity loss rate $\gamma=0.05 \omega_{\mathrm{T}}$ as considered in Figures~\ref{fig:fig4}c and ~\ref{fig:fig5}c, the quality factor $Q$ is about $20$, which provides a balance between spin cooling and loss to make experimental realization more feasible ~\cite{Shandilya2019}. Overall, negligible spin dephasing, weak spin damping and appropriate cavity loss are suggested in realizing transient DTC order in such an optomechanical system.

%%%%%%%%%%%%%%%%%%%%%%%%%%%%%%%%%%%%%%%%%%
\section{Conclusions}\label{sec:sec5}
In summary, we have investigated DTC order in a Floquet open system composed of $N$ qubits trapped in a (mechanical) cavity. The influences of all-to-all spin interactions, spin damping, spin dephasing as well as cavity loss are explored both in the thermodynamic limit and in the deep quantum regime. It is shown that the existence of dephasing will destroy the coherence of the system and finally leads the system to its trivial steady state. Without dephasing and all-to-all spin coupling, different stroboscopic dynamics in different dissipation regimes is demonstrated. First, with weak spin damping and weak all-to-all coupling, eternal DTC oscillations are observed and robust against imperfection. As the spin damping rate increases, the stroboscopic dynamics evolves irregularly accompanied by the trajectory of the scaled angular momentum vector scattered on the Bloch sphere. However, with enhancement of cavity loss, the dynamics will become more regularly and the eternal eternal DTC order will reemerge at strong cavity loss.
Besides, by growing the all-to-all coupling, we demonstrate that stroboscopic oscillations are gradually destroyed in the weak damping regime. It is interesting to show that the DTC order may be rebuilt by appropriate all-to-all coupling in the moderate damping regime.

We also focus on the few-atom cases,  the so-called deep quantum regime, the model of which describes a suspended hBN monolayer membrane with a few spin defects under a microwave drive and a Floquet magnetic field. A transient DTC lasting much longer than the decay time can be found in both weak and moderate dissipation regimes when there is no spin dephasing. Nonetheless, the existence of dephasing will destroy transient oscillations and leads the system fast to a trivial steady state, which is consistent with the results obtained by semiclassical approximation in the thermodynamic limit. We also find that stroboscopic oscillations may be more robust against the spin dephasing by increasing the spin number. Finally, the  parameters in the experimental aspect are briefly discussed and how to realizing transient DTC order in such an hBN optomechanical system is suggested.

%%%%%%%%%%%%%%%%%%%%%%%%%%%%%%%%%%%%%%%%%%

T.L. acknowledges the support from NSF (Grant No. PHY-2110591). Z.H. acknowledges the support from the Fundamental Research Funds for the Central Universities (Grant No. JUSRP21935) and the China Scholarship Council (CSC).

%\end{paracol}

%\reftitle{References}


\begin{thebibliography}{99}
	\bibitem {Abanin2019}  
	Abanin, D.A.; Altman, E.;  Bloch, I.;  Serbyn, M. Colloquium: many-body localization, thermalization, and entanglement. \emph{Rev. Mod. Phys.} \textbf{2019}, \emph{91}, 021001.
	
	\bibitem {Lyu2020} 
	Lyu, C.; Choudhury, S.; Lv, C.; Yan, Y.; Zhou, Q. Eternal discrete time crystal beating the Heisenberg limit. \emph{Phys. Rev. Res.} \textbf{2020}, \emph{2}, 033070.
	
	\bibitem {Sacha2015} 
	Sacha, K. Modeling spontaneous breaking of time-translation symmetry. \emph{Phys. Rev. A} \textbf{2015}, \emph{91}, 033617.
	
	\bibitem {Else2016} 
	Else, D.V.; Bauer, B.; Nayak, C. Floquet time crystals. \emph{Phys. Rev. Lett.} \textbf{2016}, \emph{117}, 090402.
	
	\bibitem {Khemani2016} 
	Khemani, V.; Lazarides, A.; Moessner, R.; Sondhi, S.L. Phase structure of driven quantum systems. \emph{Phys. Rev. Lett.} \textbf{2016}, \emph{116}, 250401.
	
	
	
	\bibitem {Wilczek2012} 
	Wilczek, F. Quantum time crystals. \emph{Phys. Rev. Lett.} \textbf{2012}, \emph{109}, 160401.
	
	\bibitem {Li2012} 
	Li, T.; Gong, Z.-X.; Yin, Z.-Q.; Quan, H.T.; Yin, X.; Zhang, P.; Duan, L.-M.; Zhang, X. Space-time crystals of trapped ions. \emph{Phys. Rev. Lett.} \textbf{2012}, \emph{109}, 163001.
	
	\bibitem {Huang2018} 
	Huang, Y.; Li, T.;  Yin, Z.-Q. Symmetry-breaking dynamics of the finite-size Lipkin-Meshkov-Glick model near ground state. \emph{Phys. Rev. A} \textbf{2018}, \emph{97}, 012115.
	
	\bibitem {Huang2020} 
	Huang, Y.; Guo, Q.; Xiong, A.; Li, T.;  Yin, Z.-Q. Classical and quantum time crystals in a levitated nanoparticle without drive. \emph{Phys. Rev. A} \textbf{2020}, \emph{102}, 023113.
	
	\bibitem {Sacha2020}
	Sacha, K. \emph{Time Crystals}; Springer International Publishing: Cham, Switzerland, 2020.
	
	\bibitem {Choi2017} 
	Choi, S.; Choi, J.; Landig, R.; Kucsko, G.; Zhou, H.; Isoya, J.; Jelezko, F.; Onoda, S.; Sumiya, H.; Khemani, V.; et al. Observation of discrete time-crystalline order in a disordered dipolar many-body system. \emph{Nature} \textbf{2017}, \emph{543}, 221.
	
	\bibitem {Zhang2017} 
	Zhang, J.; Hess, P.W.; Kyprianidis, A.; Becker, P.; Lee, A.; Smith, J.; Pagano, G.; I.-D. Potirniche, I.-D.;  Potter, A.C.; Vishwanath, A.; Yao, N.Y.; Monroe, C.  Observation of a discrete time crystal. \emph{Nature} \textbf{2017}, \emph{543}, 217.
	
	\bibitem {Randall2021} 
	Randall, J.; Bradley, C.E.; van der Gronden, F.V.; Galicia, A; Abobeih, M.H.; Markham, M.; Twitchen, D.J.; Machado, F.; Yao, N.Y.; Taminiau, T.H.   Many-body–localized discrete time crystal with a programmable spin-based quantum simulator. \emph{Science} \textbf{2021}, \emph{374}, 1474.
	
	\bibitem {Kyprianidis2021} 
	Kyprianidis, A.; Machado, F.; Morong, W.; Becker, P.; Collins, K.S.;  Elsel, D.V.; Feng, L.;  Hess, P.W.; Nayak, C.; Pagano, G.; Yao, N.Y.;  Monroe, C.   Observation of a prethermal discrete time crystal. \emph{Science} \textbf{2021}, \emph{372}, 1192.
	
	\bibitem {von Keyserlingk2016}
	von Keyserlingk, C.W.;  Khemani, V.;  Sondhi, S.L. Absolute stability and spatiotemporal long-range order in Floquet systems. \emph{Phys. Rev. B} \textbf{2016}, \emph{94}, 085112.
	
	\bibitem {Yao2017} 
	Yao, N.Y.; Potter, A.C.; Potirniche, I.-D.; Vishwanath, A. Discrete time crystals: rigidity, criticality, and realizations. \emph{Phys. Rev. Lett.} \textbf{2017}, \emph{118}, 030401.
	
	\bibitem {Lazarides2017}  
	Lazarides, A.; Moessner, R. Fate of a discrete time crystal in an open system. \emph{Phys. Rev. B} \textbf{2017}, \emph{95}, 195135.
	
	\bibitem {Else2017} 
	Else, D.V.; Bauer, B.; Nayak, C. Prethermal phases of matter protected by time-translation symmetry. \emph{Phys. Rev. X} \textbf{2017}, \emph{7}, 011026.
	
	\bibitem {Gong2018} 
	Gong, Z.; Hamazaki, R.; Ueda, M. Discrete time-crystalline order in cavity and circuit QED systems. \emph{Phys. Rev. Lett.} \textbf{2018}, \emph{120}, 040404.
	
	\bibitem {Zhu2019} 
	Zhu, B.;  Marino, J.; Yao, N.Y.; Lukin, M.D.; Demler, E.A. Dicke time crystals in driven-dissipative quantum many-body systems. \emph{New J. Phys.} \textbf{2019},\emph{21}, 073028.
	
	\bibitem {KL2021} 
	Ke\ss{}ler, H.; Kongkhambut, P.; Georges, C.; Mathey, M.; Cosme, J.G.; Hemmerich, A. Observation of a dissipative time crystal. \emph{Phys. Rev. Lett.} \textbf{2021}, \emph{127}, 043602.
	
	\bibitem {Lazarides2020} 
	Lazarides, A.; Roy, S.; Piazza, F.; Moessner, R. Time crystallinity in dissipative Floquet systems. \emph{Phys. Rev. Res.} \textbf{2020}, \emph{2}, 022002.
	
	\bibitem {Riera-Campeny2020}
	Riera-Campeny, A.; Moreno-Cardoner, M.; Sanpera, A. Time crystallinity in open quantum systems. \emph{Quantum} \textbf{2020}, \emph{4}, 270.
	
	
	\bibitem {Fabre1994} 
	Fabre, C.; Pinard, M.; Bourzeix, S.; Heidmann, A.;  Giacobino, E.;  Reynaud, S. Quantum-noise reduction using a cavity with a movable mirror. \emph{Phys. Rev. A} \textbf{1994}, \emph{49}, 1337.
	
	\bibitem {Mancini1994} 
	Mancini, S.; Tombesi, P. Quantum noise reduction by radiation pressure. \emph{Phys. Rev. A} \textbf{1994}, \emph{49}, 4055.
	
	\bibitem {RMP2014} 
	Aspelmeyer, M.; Kippenberg, T.J.;  Marquardt, F. Cavity optomechanics. \emph{Rev. Mod. Phys.} \textbf{2014}, \emph{86}, 1391.
	
	\bibitem {Yin2015} 
	Yin, Z.; Zhao, N.; Li, T. Hybrid opto-mechanical systems with nitrogen-vacancy centers. \emph{Sci. China Phys. Mech. Astron.} \textbf{2015}, \emph{58}, 1.
	
	\bibitem {Xu2021casimir} 
	Xu, Z.; Gao, X.; Bang, J.; Jacob, Z.; Li, T. Non-reciprocal energy transfer through the Casimir effect. \emph{Nat. Nanotechnol.} \textbf{2021}, https://doi.org/10.1038/s41565-021-01026-8.
	
	\bibitem {Chan2011} 
	Chan, J.; Alegre, T.P.M.; Safavi-Naeini, A.H.; Hill, J.T.; Krause, A.; Gr{\"o}blacher, S.; Aspelmeyer, M.; Painter, O. Laser cooling of a nanomechanical oscillator into its quantum ground state. \emph{Nature} \textbf{2011}, \emph{478}, 89.
	
	\bibitem {Liu2013} 
	Liu, Y.-c.; Hu, Y.-W.; Wong, C.W.; Xiao, Y.F. Review of cavity optomechanical cooling. \emph{Chin. Phys. B} \textbf{2013}, \emph{22}, 114213.
	
	\bibitem {Xiong2017} 
	Xiong, H.; Liu, Z. X.; Wu, Y. Highly sensitive optical sensor for precision measurement of electrical charges based on optomechanically induced difference-sideband generation. \emph{Opt. Lett.} \textbf{2017}, \emph{42}, 3630.
	
	\bibitem {Liu2017} 
	Liu, Z.-X.; Wang, B.; Kong, C.; Si, L.-G.; Xiong, H.; Wu, Y. A proposed method to measure weak magnetic field based on a hybrid optomechanical system. \emph{Sci. Rep.} \textbf{2017}, \emph{7}, 12521.
	
	\bibitem {Krause2012} 
	Krause, A. G.; Winger, M.; Blasius, T. D.; Lin, Q.; Painter, O. A high-resolution microchip optomechanical accelerometer. \emph{Nat. Photonics} \textbf{2012}, \emph{6}, 768.
	
	\bibitem {Ahn2020} 
	Ahn, J.; Xu, Z.; Bang, J.; Ju, P.; Gao, X.; Li, T. Ultrasensitive torque detection with an optically levitated nanorotor. \emph{Nat. Nanotechnol.} \textbf{2020}, \emph{15}, 89.
	
	\bibitem {LiBB2021} 
	Li, B.-B.; Ou, L.; Lei, Y.; Liu, Y.-C. Cavity optomechanical sensing. \emph{Nanophotonics} \textbf{2021}, \emph{10}, 2799.
	
	\bibitem {Stannigel2010} 
	Stannigel, K.; Rabl, P.;  S{\o}rensen, A. S.;  Zoller, P.; Lukin, M. D. Optomechanical transducers for long-distance quantum communication. \emph{Phys. Rev. Lett. } \textbf{2010}, \emph{105}, 220501.
	
	\bibitem {Stannigel2012} 
	Stannigel, K.;   Komar, P.;  Habraken, S. J. M.; Bennett, S. D.;  Lukin, M. D.; Zoller, P.;  Rabl, P. Optomechanical quantum information processing with photons and phonons. \emph{Phys. Rev. Lett. } \textbf{2012}, \emph{109}, 013603.
	
	\bibitem {Karg2020} 
	Karg, T. M.;  Gouraud, B.;  Ngai, C. T.; Schmid, G.-L.;  Hammerer, K.; Treutlein, P. Light-mediated strong coupling between a mechanical oscillator and atomic spins 1 meter apart. \emph{Science} \textbf{2020}, \emph{369}, 174.
	
	\bibitem {Frisk Kockum2019}
	Frisk Kockum, A.;  Miranowicz, A.; De Liberato, S.; Savasta, S.; Nori, F. Ultrastrong coupling between light and matter. \emph{Nat. Rev. Phys.} \textbf{2019}, \emph{1}, 19.
	
	\bibitem {Akinwande2017} 
	Akinwande, D.; Brennan, C.J.; Bunch, J.S.; Egberts, P.; Felts, J.R.; Gao, H.; Huang, R.; Kim, J.-S.; Li, T.; Li, Y.; et al. A Review on mechanics and mechanical properties of 2D materials--graphene and beyond. \emph{Extreme Mech. Lett.} \textbf{2017}, \emph{13}, 42.
	
	\bibitem {Tran2016}  
	Tran, T. T.; Bray, K.; Ford, M. J.; Toth, M.;  Aharonovich, I. Quantum emission from hexagonal boron nitride monolayers. \emph{Nat. Nanotechnol.} \textbf{2016}, \emph{11}, 37.
	
	
	\bibitem {Liu2019} 
	Liu, X.; Hersam, M. C. 2D materials for quantum information science. \emph{Nat. Rev. Mater.} \textbf{2019}, \emph{4}, 669.
	
	\bibitem {Xia2014} 
	Xia, F.;  Wang, H.;  Xiao, D.;  Dubey, M.;  Ramasubramaniam, A. Two-dimensional material nanophotonics. \emph{Nat. Photonics} \textbf{2014}, \emph{8}, 899.
	
	\bibitem {Klusek2010}
	S{\l}awi{\'n}ska, J.;  Zasada, I.; Kosi{\'n}ski, P.; Klusek, Z. Reversible modifications of linear dispersion: graphene between boron nitride monolayers. \emph{Phys. Rev. B} \textbf{2010}, \emph{82}, 085431.
	
	\bibitem {Cadiz2018}
	Cadiz, F.;Robert, C.; Courtade, E.; Manca, M.; Martinelli, L.; Taniguchi, T.; Watanabe, K.; Amand, T.; Rowe, A.C.H.;
	Paget, D.; et al. Exciton diffusion in WSe2 monolayers embedded in a van der Waals heterostructure. \emph{Appl. Phys. Lett.} \textbf{2018}, \emph{112}, 152106.
	
	
	\bibitem {Tran2017}
	Tran, T. T.; Wang, D.; Xu, Z.-Q.; Yang, A.; Toth, M.; Odom, T. W.; Aharonovich, I. Deterministic coupling of quantum emitters in 2D materials to plasmonic nanocavity arrays. \emph{Nano Lett.} \textbf{2017}, \emph{17}, 2634.
	
	\bibitem {Caldwell2019}
	Caldwell, J. D.; Aharonovich, I.; Cassabois, G.; Edgar, J. H.; Gil, B.; Basov, D. Photonics with hexagonal boron nitride. \emph{Nat. Rev. Mater.} \textbf{2019}, \emph{4} 552.
	
	\bibitem {Gao2020}
	Gao, X.; Yin, Z.; Li, T. High-speed quantum transducer with a single-photon emitter in a 2D resonator. \emph{Annalen Physik} \textbf{2020}, \emph{532}, 2000233.
	
	\bibitem {Wu2021}
	Wu, K.; Zhang, H. Chen, Y.; Luo, K.; Xu, K. All-silicon microdisplay using efficient hot-carrier electroluminescence in standard 0.18 $\mu$m CMOS technology. \emph{IEEE Electron Device Lett.} \textbf{2021}, \emph{42}, 541.
	
	\bibitem {Xu2021}
	Xu, K.; Silicon electro-optic micro-modulator fabricated in standard CMOS technology as components for all silicon monolithic integrated optoelectronic systems. \emph{J. Micromech. Microeng.} \textbf{2021}, \emph{31}, 054001.
	
	\bibitem {Kim2018}  
	Kim, S.;  Fr\"{o}ch, J. E.;  Christian, J.;  Straw, M.;  Bishop, J.; Totonjian, D.;  Watanabe, K.;  Taniguchi, T.;  Toth, M.;  Aharonovich, I. Photonic crystal cavities from hexagonal boron nitride. \emph{Nat. Commun.} \textbf{2018}, \emph{9}, 2623.
	
	
	\bibitem {Shandilya2019} 
	Shandilya, P. K.; Fr\"{o}ch, J. E.; Mitchell, M.; Lake, D. P.;  Kim, S.;  Toth, M.; Behera, B.;  Healey, C.; Aharonovich, I.   Barclay, P. E. Hexagonal boron nitride cavity optomechanics. \emph{Nano Lett.} \textbf{2019}, \emph{19}, 1343.
	
	\bibitem {Abdi2017} 
	Abdi, M.;  Hwang, M.-J.; Aghtar, M.; Plenio, M. B.; Spin-mechanical scheme with color centers in hexagonal boron nitride membranes. \emph{Phys. Rev. Lett.} \textbf{2017}, \emph{119}, 233602.
	
	\bibitem {Abdi2019}  
	Abdi, M.; Plenio, M.B. Quantum effects in a mechanically modulated single-photon emitter. \emph{Phys. Rev. Lett.} \textbf{2019}, \emph{122}, 023602.
	
	\bibitem {Gottscholl2020}  
	Gottscholl, A.; Kianinia, M.; Soltamov, V.; Orlinskii, S.; Mamin, G.; Bradac, C.; Kasper, C.;  Krambrock, K.; Sperlich, A.; Toth, M.; Aharonovich, I. Initialization and read-out of intrinsic spin defects in a van der waals crystal at room temperature. \emph{Nat. Mater.} \textbf{2020}, \emph{19}, 540.
	
	\bibitem {Chejanovsky2021}  
	Chejanovsky, N.; Mukherjee, A.; Geng, J.; Chen, Y.-C.; Kim, Y.; Denisenko, A.;  Finkler, A.; Taniguchi, T.; Watanabe, K.; Dasari, D.B.R.; et al. Single-spin resonance in a van der Waals embedded paramagnetic defect. \emph{Nat. Mater.} \textbf{2021}, \emph{20}, 1079.
	
	\bibitem {Gao2021}  
	Gao, X.; Jiang, B.; Allcca, A. E. L.; Shen, K.; Sadi, M. A.; Solanki, A. B.; Ju, P.; Xu, Z.; Upadhyaya, P.; Chen, Y. P.; Bhave, S. A.; Li, T. High-contrast plasmonic-enhanced shallow spin defects in hexagonal Boron Nitride for quantum sensing. \emph{Nano Lett.} \textbf{2021}, \emph{21}, 7708.
	
	\bibitem {LiB2020}  
	Li, B.; Li, X.; Li, P.; Li, T. Preparing squeezed spin states in a spin–mechanical hybrid system with silicon-vacancy centers. \emph{Adv. Quantum Technol.} \textbf{2020}, \emph{3}, 2000034.
	
	
	\bibitem {Morrison2008}  
	Morrison, S.; Parkins, A.S. Dynamical quantum phase transitions in the dissipative Lipkin-Meshkov-Glick model with proposed realization in optical cavity QED. \emph{Phys. Rev. Lett.} \textbf{2008}, \emph{100}, 040403.
	
	\bibitem {Russomanno2017} 
	Russomanno, A.; Iemini, F.; Dalmonte, M.; Fazio, R. Floquet time crystal in the Lipkin-Meshkov-Glick model. \emph{Phys. Rev. B} \textbf{2017}, \emph{95}, 214307.
\end{thebibliography}
\end{document}